# Generation of Broadband Mid-IR and UV Light in Gas-Filled Single-Ring Hollow-Core PCF

**MARCO CASSATARO,**[1,*] **DAVID NOVOA,**[1] **MEHMET C. GÜNENDI,**[1] **NITIN N. EDAVALATH,**[1] **MICHAEL H. FROSZ,**[1] **JOHN C. TRAVERS,**[2] **AND PHILIP ST.J. RUSSELL**[1]

[1]*Max Planck Institute for the Science of Light, Staudtstrasse 2, 91058 Erlangen, Germany*
[2]*School of Engineering and Physical Sciences, Heriot-Watt University, Edinburgh EH14 4AS, U.K.*
*\*marco.cassataro@mpl.mpg.de*

**Abstract:** We report generation of an ultrafast supercontinuum extending into the mid-infrared in gas-filled single-ring hollow-core photonic crystal fiber (SR-PCF) pumped by 1.7 µm light from an optical parametric amplifier. The simple fiber structure offers shallow dispersion and flat transmission in the near and mid-infrared, enabling the generation of broadband spectra extending from 300 nm to 3.1 µm, with a total energy of a few µJ. In addition, we report the emission of ultraviolet dispersive waves whose frequency can be tuned simply by adjusting the pump wavelength. SR-PCF also provides an effective means of compressing and delivering tunable ultrafast pulses in the near and mid-infrared spectral regions.

## 1. Introduction

Broad-band supercontinuum (SC) sources have multiple applications in, for example, spectroscopy [1], synthesis of ultrashort pulses [2], optical coherence tomography [3] and microscopy [4]. SC generation received a significant boost with the advent of photonic crystal fiber (PCF), which offers long well-controlled diffraction-free path-lengths—a significant advance over bulk or free-space systems [5]. In addition, PCF enables the dispersion to be engineered over a wide range, for example permitting anomalous dispersion at near-infrared (NIR) wavelengths, which allows soliton dynamics to be used in the generation of ultrabroad supercontinua [5].

While silica-based solid-core PCFs are convenient for SC generation in the NIR and visible regions, it is not possible to further extend the spectrum into deep ultraviolet (DUV) due to color-center-related damage, nor into the mid-infrared (MIR) because of strongly increasing absorption. Materials with wider transmission windows, such as fluoride glass, in principle overcome these limitations [6, 7], but at the cost of mechanical fragility and more complex fabrication. In addition, the generated SC energy is limited by the damage threshold, which for solid materials is much lower than for gases.

Generating high-energy supercontinua that extend into the MIR is of great interest, because this spectral region is important in spectroscopy [8] and biomedicine [9]. Gas-filled hollow-core PCF (HC-PCF) provides an ideal alternative to solid-core fibers, since it offers a much higher damage threshold as well as negligible absorption (by suitable choice of gas). Furthermore, it allows easy and precise pressure-tuning of the nonlinearity and dispersion, enabling precise control of the dynamics of SC generation [10]. Kagomé-style HC-PCFs, in particular, have been successfully used to generate broad supercontinua [10]. More recently single-ring HC-PCFs (SR-PCFs), with a simpler transverse structure compared to kagomé HC-PCFs, have been shown to provide excellent transmission at longer wavelengths [11], making them ideal for the generation of MIR supercontinua. In this paper we report the generation of a broad SC spanning more than 3 octaves from the DUV to the MIR, with a total energy of a few µJ. Ultraviolet (UV) light is generated by dispersive wave emission from self-compressed pump pulses, and is wavelength tunable by tuning the pump wavelength, pressure and pulse energy.

## 2. Single-Ring Hollow-Core Photonic-Crystal Fiber

Fig. 1(a) shows a scanning electron micrograph of the transverse structure of the SR-PCF used in this work. It consists of a hollow core surrounded by six non-touching silica capillaries arranged within a hexagonal jacket. The core diameter is $D \sim 51$ µm, the diameter of the hollow capillaries is $d \sim 21$ µm, and the thickness of the capillary walls is ~280 nm. Strong phase-mismatch between the light in the core and the modes supported by the hollow capillaries prevents the core radiation from leaking into the cladding, thus allowing the fiber to guide through anti-resonant reflection of core light. While the conditions for anti-resonant guidance are satisfied for the fundamental core-mode in a broad wavelength range, except for a few well-defined resonances of the capillary walls, it has been recently shown that for $d/D \sim 0.68$ the higher-order $LP_{11}$-like mode of the core is phase-velocity matched to the fundamental mode of the cladding capillaries, making the hollow-core fiber effectively *endlessly* (for all guided wavelengths) single mode (hESM) [12]. Since the fiber in Fig. 1(a) has $d/D \sim 0.41$, far from the hESM condition, we expect HOMs to play a role in the nonlinear dynamics, as we shall see below.

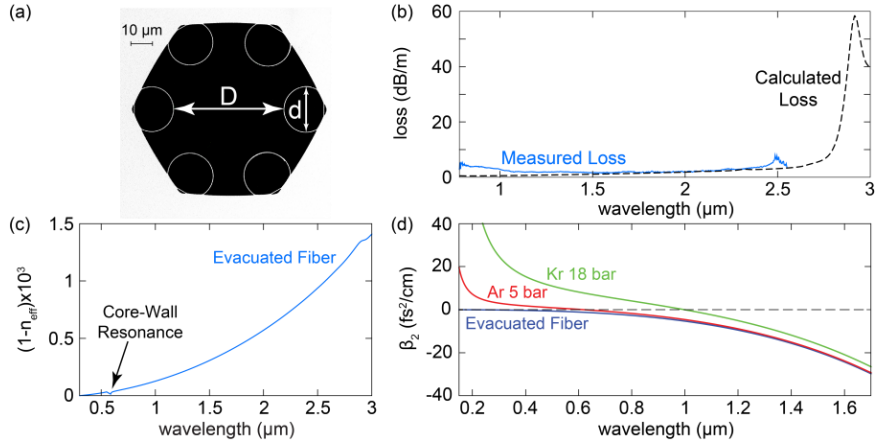

Fig. 1. (a) Scanning electron micrograph of the transverse structure of the SR-PCF. (b) Measured (solid blue line) and simulated (dashed black line) loss of the SR-PCF. The FEM only accounts for confinement loss, i.e., the attenuation of silica is not included. (c) Calculated effective index for the $LP_{01}$-like core mode. It has a very smooth profile following the capillary model, except for a small, localized resonance at ~550 nm. (d) Group-velocity dispersion parameter $\beta_2$ calculated analytically through the capillary model, showing the tunability of the dispersive properties of the system.

In contrast to kagomé-style HC-PCFs, the simpler structure of SR-PCF leads to a flat transmission spectrum, with just a few loss bands whose wavelengths are straightforwardly predictable [12]. Fig. 1(b) shows that the loss of the fiber is limited to a few dB/m throughout the NIR, up to ~2.4 µm (limit of the broadband source used to measure the fiber attenuation). Although this loss is negligible for our work—the dynamics observed takes place in only a few centimeters of HC-PCF—it has been already shown that SR-PCFs can provide much lower attenuation [13]. Finite element modeling (FEM), represented by the dashed-black curve in Fig. 1(b), predicts the confinement loss to increase dramatically beyond 2.7 µm. These calculations neglect the absorption of fused silica in the MIR, so that the actual loss is expected to be higher at longer wavelengths when the light-glass overlap increases. As a result, we would expect the spectrum reported in this work (see below) to extend further into the MIR if lower loss SR-PCFs were used, for example by using a larger core diameter. The almost loss-band-free transmission results in a very smooth wavelength-dependent modal refractive index, as shown in Fig. 1(c) where the calculated effective index of the $LP_{01}$-like core mode is plotted versus wavelength (the small fluctuation at ~550 nm is caused by phase-matching to a resonance in the capillary walls). The shallow profile of the effective index [12] makes the system ideal for exploiting soliton dynamics and generating a SC.

As in the case of kagomé-style HC-PCFs, the weakly anomalous geometrical dispersion can be balanced by the normal dispersion of the filling gas [10]. The group-velocity dispersion $\beta_2$ can be readily calculated using a simple analytical model for wide-bore capillaries, well suited to SR-PCFs [12]. The zero-dispersion-wavelength (ZDW) is tunable from the visible to the NIR, and as a consequence soliton dynamics is enabled when a NIR or MIR pulse is launched inside the fiber, enabling SC generation and emission of resonant radiation in the form of dispersive waves, as will be discussed in the following section.

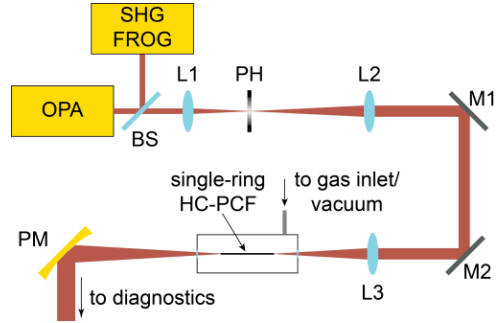

Fig. 2. Experimental set-up. The pump pulses were generated by an optical parametric amplifier, and are temporally characterized using a home-built SHG-FROG. The pulses were spatially filtered and expanded through the focusing lens L1, the collimating lens L2 (both made of $CaF_2$) and a pin-hole (PH) with a 100 µm aperture. The beam is then focused through the $CaF_2$ lens L3 into the SR-PCF, placed inside a gas cell connected to a vacuum pump and a gas line. The pump pulses enter the gas cell through a $CaF_2$ window. The divergent beam at the fiber output exits the gas cell through another $CaF_2$ window, and is then collimated by a parabolic mirror (PM) and sent to the diagnostic systems.

## 3. Experimental results

The experimental set-up is shown in Fig. 2. As pump pulses we used the idler signal of an optical parametric amplifier, tunable between 1.6 and 2.3 µm, with energy up to 100 µJ, repetition rate 1 kHz and almost transform-limited duration of 28.5 fs (measured with a home-built second-harmonic generation frequency-resolved optical gating system). These pulses were launched into the SR-PCF described in Section 2 after passing through a spatial filter. Coupling efficiencies up to 90% were achieved for low-energy (< 1 µJ) pulses; for higher energies the coupling efficiency decreases because of nonlinear absorption and photo-ionization of the gas. The SR-PCF was enclosed in a gas cell, which could be either evacuated or filled with gas. The output beam was collimated and diverted to diagnostics for spectral and beam profile characterization.

The upper plot in Fig. 3(a) shows the results when 10 µJ-pulses centered at 1.7 µm were launched into a 5-cm-long SR-PCF filled with krypton at 18 bar. With these parameters, the ZDW is ~1045 nm, thus placing the pump pulses in the anomalous dispersion regime. The soliton order is 13, below the threshold for modulational instability, rendering the soliton dynamics coherent [5]. The NIR-MIR spectrum (solid-red line) was measured using a semiconductor-detector-based monochromator (not intensity calibrated), while the spectrum in the visible-UV was characterized with a calibrated CCD-based spectrometer (solid-blue line). The output spectrum for low-energy pumping (~500 nJ) is shown for comparison (solid-black line). All the spectra are scaled so as to account for the wavelength-dependent divergence of the beam leaving the fiber. The measured spectrum shows that in-fiber soliton dynamics generates a SC covering more than 3 octaves from the DUV to the MIR, the 30 dB limits being at 270 nm on the blue side, and 3.1 µm on the red side. The total energy contained in the SC is 4 µJ, measured with a thermal power sensor. Note that the spectrum is continuous between the DUV and MIR and well above the noise floor of the spectral measurements (corresponding to –45 dB for the visible-UV and –37 dB for the MIR).

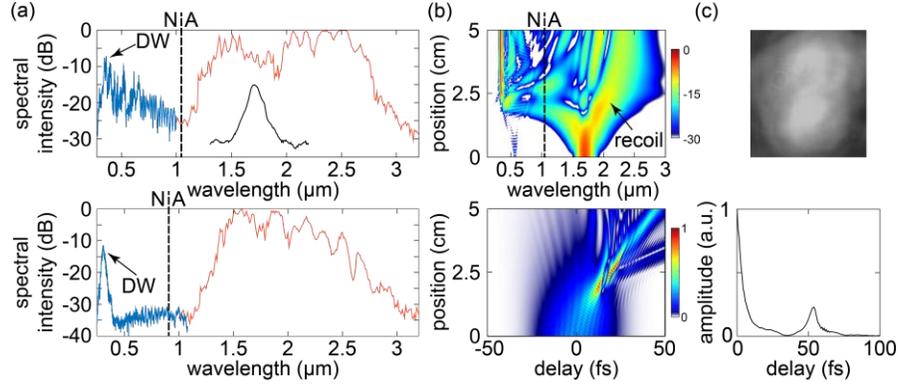

Fig. 3. (a) The red curves show the spectrum measured by a monochromator in the near/mid-infrared, and the blue curves by a CCD-based spectrometer in the visible/UV. The black dashed lines mark the ZDW of the system, and N and A denote normal and anomalous dispersion. The positions of the dispersive wave peaks are marked by the initials DW. Upper plot: pump pulse energy 10 µJ, Kr pressure 18 bar. The black curve shows the spectral intensity (normalized to the maximum of the red spectrum)) generated with low-energy pump pulses (~500 nJ). Lower plot: pulse energy 4 µJ and Kr pressure 10 bar. (b) Simulations of the spectral (upper) and temporal (lower) evolution of a Gaussian pulse with an energy of 5 µJ propagating along a 5 cm long SR-PCF filled with krypton at 18 bar. The dashed line in the spectral plot indicates the ZDW. (c) Upper: Far-field profile of the supercontinuum beam filtered at 490 nm and imaged onto a CCD camera. Lower: Fourier transform of the spectral intensity in the lower part of (a).

Two main features are evident in the measured spectra. The first is a band of resonant radiation in the DUV, peaking at 347 nm, caused by phase-matching of the self-compressed broadband pump soliton to UV light in the normal dispersion regime. The spectral position of the dispersive wave agrees well with the predictions of the nonlinear phase-matching condition [14]:

$$\beta(\omega) - \beta(\omega_{sol}) - \beta_1(\omega_{sol})[\omega - \omega_{sol}] - \gamma P_C/2 = 0 \qquad (1)$$

where $\beta$ is the linear propagation constant, $\omega_{sol}$ the central angular frequency of the soliton, $\beta_1$ the inverse of the group velocity, $\gamma$ the nonlinear fiber parameter and $P_C$ the peak power of the soliton at the maximum compression point. A second clear feature is the spectral recoil of the pump: consequent to the emission of resonant radiation in the UV region, the soliton wavelength red-shifts so as to conserve the total energy. This frequency-downshift, which has physical origin entirely different from the Raman self-frequency shift [15], is adiabatic owing to the smooth dispersion landscape.

To illustrate how the system dynamics can be engineered for a given application, a second measurement (Fig. 3(a), lower) was performed at a pressure of 10 bar and a lower input energy (4 µJ, corresponding to soliton order 8). The lower soliton order leads to smoother self-compression, and although the supercontinuum has lower spectral intensity in the visible (though still above the noise-floor of the CCD-based spectrometer) and extends less far into the MIR (the 30 dB limit being at ~2.9 µm), a more distinct dispersive wave band appears at 310 nm.

An additional feature that is evident in the measurements is a quasi-periodic modulation of the spectral intensity. We attribute this to inter-modal beating between the fundamental and low loss higher-order modes (especially the $LP_{11}$-like mode)—a consequence of the sub-optimal $d/D$ ratio [12]. This is supported by inspection of the output mode profile (Fig. 3(c), upper), imaged in the far-field with a CCD camera after filtering out a 10 nm-wide band, centered at 490 nm, from the SC shown in the lower plot in Fig. 3(a). It is clear that the profile is a mixture of $LP_{01}$ and $LP_{11}$-like modes. To further check this, the lower plot in Fig. 3(c) shows the Fourier transform of the SC spectral intensity [16], from which we see that the interference period observed on the spectrum corresponds to a delay of 53 fs, which is the separation between the two modes at the pump wavelength after ~2.5 cm of co-propagation inside the fiber.

In Fig. 3(b) we show the spectral (upper) and temporal (lower) evolution of the pulse as it propagates along the fiber, simulated using the unidirectional field equation [17]. The parameters used in the simulations are the same as in the experiment shown in the upper plot in Fig. 3(a), except that the input Gaussian pulse energy was set to 5 µJ (10 µJ in the experiment) to compensate for nonlinear absorption in the gas not taken into account in the simulations. The modeling shows that the pump pulse undergoes an initial stage of self-compression during which its spectrum massively broadens, followed by generation of a UV dispersive wave. The simulations included only the $LP_{01}$-like core-mode, so that the modulation observed in the measured spectrum is not present, providing further confirmation of its origin in excitation of HOMs. The simulated spectral evolution shows qualitative agreement with the measured spectrum, reproducing all the main features observed in Fig. 3(a), i.e., the presence of a dispersive wave peaking at ~347 nm, the spectral recoil that red-shifts the soliton as it propagates, and the extension of the SC out to ~3.1 µm. As discussed in section 2, we attribute the long-wavelength cut-off of the SC spectrum to confinement loss in the fiber, which becomes very significant from 2.7 µm onwards.

These results demonstrate that the system can generate high-energy, ultra-wide supercontinua in the MIR wavelength range. Compared to other successful schemes that have been proposed [18], it has the advantage of easy implementation, without need for coherent recombination. As the SC spectrum is expected to be fully coherent, it would be possible to compress it using a suitable dispersive element and obtain nearly single-cycle pulses in the MIR spectral region [2]. For applications that require the SC to be in a pure fundamental mode, the use of a SR-PCF with a suitable $d/D$ ratio would allow for complete elimination of the HOM contributions observed here [12].

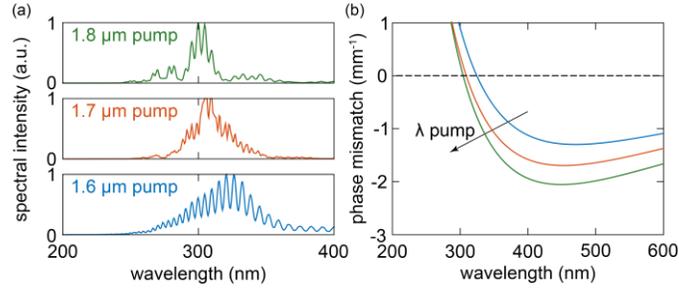

Fig. 4. (a) Tuning the wavelength of the dispersive wave by varying the pump wavelength. Pulses of wavelength 1.6 (lower plot), 1.7 (middle plot) and 1.8 µm (upper plot) are coupled into the single-ring HC-PCF filled with krypton at 10 bar. The generated dispersive waves peak at 322, 310 and 305 nm, respectively. (b) Phase-mismatch between linear waves and the pump soliton for the same parameters considered in (a).

## 4. Wavelength-tunability of the system

The flat, broad transmission profile of the SR-PCF makes it very versatile. For example, it is possible to tune the position of the emitted dispersive waves just by changing the input wavelength of the pulses launched into the fiber: this means that pump pulses with different central wavelengths can be compressed to single-cycle durations using the same SR-PCF, something never demonstrated in other broadband-guiding PCFs such as kagomé-style HC-PCF.

Fig. 4(a) displays the dispersive wave spectra emitted in the SR-PCF filled with 10 bar of krypton (ZDW at 820 nm) when it is pumped with ultrashort pulses of 1.6, 1.7 and 1.8 µm central wavelengths. The energy of the input pulses was chosen so as to yield a soliton order of ~8, resulting in clean self-compression and dispersive wave emission. As the pump wavelength shifts over 200 nm, the nonlinear dynamics change smoothly, resulting in the emission of dispersive waves tunable over 17 nm. This is predicted by the phase-matching condition Eq. (1), as can be seen in Fig. 4(b), where the phase-mismatch is plotted versus wavelength. It is well known that the idler pulses generated by a white-light seeded OPA are intrinsically CEP-stable, a property that is expected to be transferred to the entire SC, since the soliton-based broadening process is coherent [19].

As demonstrated in earlier work, the dispersive wave frequency can also be tuned by varying the gas pressure, the input pulse energy or the gas species [14]. Together with the performance of the system in the MIR, this makes SR-PCF a unique vehicle for simultaneous access of many different spectral regions, raising the prospect of novel ultrafast pump-probe arrangements.

## 5. Conclusions and outlook

Gas-filled SR-PCF can be used to generate ultrabroadband spectra extending from the UV to the MIR. The SR-PCF structure combines the advantages of kagomé HC-PCFs (pressure control of the nonlinear and dispersive properties) with a much simpler microstructure that is easier to fabricate and almost free of intrusive loss bands. The improved performance of SR-PCF in the MIR makes it possible to generate a SC more than three octaves wide, with a total energy of 4 µJ, from pump pulses of 10 µJ. The system is highly versatile, permitting the UV dispersive waves to be wavelength-tuned by changing the pump wavelength without need to alter the system. Further improvements are expected if SR-PCFs with larger core-diameters are used, so as to reduce loss in the MIR, allowing the spectrum to extend to even longer wavelengths, or designing in endlessly single-mode (hESM) properties so to obtain cleaner output beam profiles.

The spectra generated are coherent and broad enough to support few-cycle pulses. With suitable filtering, propagation through dispersion compensating materials or reflection at phase-compensating mirrors, the SC pulses can be compressed [2], making this compact, versatile system an interesting ultrafast source for spectroscopy both in the MIR molecular fingerprinting region [6], where all molecules have phononic bond resonances, and in the UV region, where single-photon excitations and electronic resonances can be addressed [20]. The versatility and compactness of the source also makes it attractive for broadband MIR seed generation for amplification in the new generation of OPCPA based field synthesizers [18].